# Quantitative predictions from competition theory with incomplete information on model parameters tested against experiments across diverse taxa

Hugo Fort[1*]

[1]Institute of Physics, Faculty of Science, Universidad de la República, Iguá 4225, Montevideo 11400 Uruguay;

* E-mail: hugo@fisica.edu.uy

**Abstract**

The capacity of community ecology for making quantitative predictions is often limited by incomplete empirical information which precludes obtaining reasonable estimates of model parameters. This is particularly the case for communities with large species richness $S$ since it is practically impossible to perform the $S$ monoculture experiments (to obtain the species carrying capacities) plus the $S \times (S-1)/2$ pairwise experiments (required to estimate the entire set of interspecific interaction coefficients of the interaction or community matrix). However quantitative predictive tools are vital for understanding the fate of ecological communities.

Here we derive an analytical approximation for predicting the relative yield total (or equivalently the mean relative yield) as a function of the mean intensity of the interspecific competition and the species richness. The rationale is that with incomplete information on model parameters it seems more reasonable to attempt to predict "macroscopic" or mean quantities, defined for the whole community of competing species, rather than making detailed "microscopic" predictions, like species abundances.

This method, with only a fraction of the model parameters (carrying capacities and competition coefficients), is able to predict accurately empirical measurements covering a wide variety of taxa −algae, plants, protozoa, etc.

Our aim with this approach is to contribute in making community ecology a more quantitative and predictive science. We argue that such parsimonious modeling is preferable than more realistic theories involving additional parameters which can be very difficult to measure.

## I. INTRODUCTION

Competition plays a key role in structuring ecological communities [1,2]. The classical theoretical framework for competition is the Lotka-Volterra competition theory (LVCT) [3,4]. The Lotka-Volterra equations of competition describe the population dynamics of $S$ species competing for some common resource in a community as:

$$\frac{1}{Y_i}\frac{dY_i}{dt} = r_i\left(1 - \frac{\sum_{j=1}^{S}\alpha_{ij}Y_j}{K_i}\right) \qquad i = 1,...,S, \qquad (1)$$

where $Y_i$ is the yield of species $i$ (either its population density, biomass density, biovol, etc.), $r_i$ is its maximum per capita growth rate, $K_i$ s the carrying capacity of species $i$ (the asymptotic value the yield reaches when the species is isolated from the other competing species) and $\alpha_{ij}$ is the coefficient of competition of species $j$ over species $i$ (the intraspecific competition coefficients of a species with itself, $\alpha_{ii}$, are customarily taken equal to 1). The LVCT can be thought as the first order or linear approximation in a Taylor series expansion of the per-capita growth rates of species about the equilibrium points of a more complex and general theory of competition [4,5].

For large values of $S$ the focus of LVCT has been almost exclusively on community stability and species coexistence, not on making quantitative predictions to be tested against experimental data, like relative species abundance in equilibrium [6]. A main difficulty for quantitative analysis of the equilibrium state is how to obtain the set of parameters $\{K_i; \alpha_{ij}\}$ of eq. (1) which are relevant at equilibrium. A straightforward procedure is to perform a) the $S$ single species or monoculture experiments, and from each of them to estimate $K_i$ as the asymptotic biomass density of species $i$; and b) the $S\times(S-1)/2$ pairwise experiments and for each of them, from the pair of the biculture yields, $B_i$ and $B_j$, infer $\alpha_{ij}$ and $\alpha_{ji}$ (see below). Among several practical limitations of this approach, the principal one is that it is feasible provided $S$ is not too large since the number of required experiments grows as $S^2$.

Here we test the LVCT as a quantitatively predictive tool for equilibrium even with an incomplete knowledge of the $S^2$ model parameters, as it usually happens in real life. Therefore we do not attempt of making detailed "microscopic" predictions, like species yields $\{Y_i\}$. Rather we focus on two global or "macroscopic" quantities, for the whole community of competing species. In order to have a robust description, independent of how the yield is measured in different communities (population density, biomass density, biovol, etc.) and capable of making comparisons among widely different taxa, these macroscopic quantities are built from relative yields $y_i = Y_i/K_i$ (*i.e.* the species yield in mixture normalized by its yield in monoculture). The first macroscopic quantity is the sum of the $S$ relative yields, or relative yield total, *RYT* :

$$RYT = \sum_{i=1}^{S} y_i . \qquad (2)$$

This index allows comparing community productivity on a relative basis. Agricultural science is an example of an area where it is widely used to determine productivity in crop mixtures [7]. A *RYT* > 1 denotes species complementarity leading thus to increased total resource use [8]. The *RYT* divided by *S* is equal to the mean relative yield *MRY*, another measure of community productivity, which can be used to assess the impact of biodiversity loss on the ability of ecosystems to provide ecological services. The other macroscopic quantity is the mean intensity of inter-specific competition, *a*. To define it we rewrite the equilibrium solutions of (1) in terms of relative yields, $y_i$, as:

$$\sum_{j=1}^{S} a_{ij} y_j = 1, \qquad i = 1,...,S, \qquad (3)$$

where $a_{ij}=\alpha_{ij}K_j/K_i$. (Notice that the diagonal terms of the matrix **A**, corresponding to intra-specific competition, remain equal to 1 since $a_{ii}=\alpha_{ii}K_i/K_i = \alpha_{ii} = 1$). Thus *a* is defined as the mean over the off-diagonal elements of **A**:

$$a = \langle a_{ij} \rangle_{i \neq j}. \qquad (4)$$

We derive an analytical approximate expression that allows to predict the *RYT* as a function of an estimate of *a* −rather than using the complete set of interspecific interaction coefficients $\{a_{ij}\}$− and of the species richness *S*. We show that this theoretical *RYT* reproduces quite accurately the empirical *RYT* obtained from experiments across diverse taxa.

Since our aim is to contribute in making community ecology a more quantitative and predictive science, we discuss possible applications of this framework.

**II. METHODS AND MATERIALS**

Solving eq. (3) we obtain in matrix form

$$\mathbf{y} = \mathbf{A}^{-1}\mathbf{1}. \qquad (5)$$

By taking the sum of the entries in both sides of (5) the *RYT* can be written as:

$$RYT = \sum_{i=1}^{S} y_i = \sum_{i=1}^{S}\sum_{j=1}^{S} (\mathbf{A})_{ij}^{-1} = S^2 \langle \mathbf{A}^{-1} \rangle. \qquad (6)$$

That is, the *RYT* is proportional to the mean over the elements of the inverse of the competition matrix $\mathbf{A}^{-1}$.

If we consider a "mean field" approximation −in the space of interaction parameters− for the competition matrix (denoted by a hat), *i.e.* all the off-diagonal interspecific interaction coefficients replaced by their mean value *a*:

$$\hat{a}_{ij} = \begin{cases} 1 & \text{if } i = j \\ a & \text{if } i \neq j \end{cases}. \tag{7}$$

An estimation of *a* can be obtained either from the statistical sampling of pairwise experiments or from knowing the intensity of resource overlap among different species in a community competing for this (these) resource(s) [9]. It is easy to show that the inverse of matrix (7) has the form:

$$\hat{a}^{-1}_{ij} = \begin{cases} \dfrac{1+(S-2)a}{1+(S-2)a-(S-1)a^2} & \text{if } i = j \\ \dfrac{-a}{1+(S-2)a-(S-1)a^2} & \text{if } i \neq j \end{cases}. \tag{8}$$

Therefore,

$$\left\langle \hat{\mathbf{A}}^{-1} \right\rangle = \frac{S[1+(S-2)a]-S(S-1)a}{1+(S-2)a-(S-1)a^2} \Big/ S^2 = \frac{1-a}{S(1+(S-2)a-(S-1)a^2)} = \frac{1}{S(1+(S-1)a)},$$

and substituting the above result into (6), we obtain the relative yield total produced by the simplified competition matrix (7) as a function of *a* and *S*:

$$\widehat{RYT} = \frac{S}{1+(S-1)a}. \tag{9}$$

Just in passing notice that

$$S^2 \left\langle \hat{\mathbf{A}}^{-1} \right\rangle = \frac{S}{1+(S-1)a} = \frac{1}{\left\langle \hat{\mathbf{A}} \right\rangle},$$

and thus the approximated *RYT* is the inverse of the mean value of $\hat{\mathbf{A}}$.

To test the approximated formula (9) for the *RYT* we searched for experiments that for a given community of *S* species, besides measuring the yields of

i) these *S* species in competition, $Y_i$ (one treatment), also measured the yields of:

ii) species in isolation or monocultures $K_i$ (*S* treatments) and of

iii) species in bicultures, $B_i$ ($S(S-1)/2$ treatments).

i) & ii) serve to compute the relative yields $y_i$ and from them, using (2), the *RYT*. From i) & iii) we compute the $b_i$ and, inverting (3) for $S = 2$, we can extract the competition coefficients for the pair of species *i-j* as

$$a_{ij} = (1-b_i)/b_j, \quad a_{ji} = (1-b_j)/b_i, \tag{10}$$

and then by taken their mean obtain *a*. There are lots of published experimental studies that measured all these quantities for $S = 2$, but the number decreases quickly as $S$ grows. In fact, for $S > 8$ we couldn't find an experiment in which the totality of the $S(S+1)/2$ treatments was carried out. Therefore we took the mean $\langle a_{ij}\rangle_{i\neq j}$ over the subset of pairwise experiments that were carried out (see next section).

The LVCT has also been criticized because predictions in general assume that the community is at equilibrium –a theoretical ideal rarely achieved in natural communities [10]– and because interspecific interactions different from competition –like facilitation or cooperation – are quite frequent in nature and spoil the applicability of the theory [11]. Therefore two additional requirement on the experiments were that:

iv) the species biomasses vs. time seemed to stabilize at an equilibrium state (and we preferably take into account the last measurement of the temporal series);

v) the dominant interspecific interaction was competition (*i.e.* we discarded those experiments in which facilitation or positive interactions between species dominated over competition).

We found in the literature a total of 77 experiments verifying the above requirements: 52 with $S = 2$ [12] (see Table S1) plus 25 with $S > 2$, ranging from $S = 3$ to $S = 32$ [13-25] (see Table 1). Experiments were completed in laboratory or in the field under natural conditions (see Table 1 and S1 for a comprehensive list of experiments used and their references).

Box 1 summarizes the definition of the different quantities we consider.

**Box 1**

| quantity | defined for | definition | symbol |
|---|---|---|---|
| Yield in monoculture | species | Biomass density, biovol, etc. of a species isolated from the other species (= carrying capacity). | $K_i$ |
| Yield in community (yield in $S$-polyculture) | species | Biomass density, biovol, etc. of a species $i$ interacting with the other $S$-1 species making up the community. | $Y_i$ |
| Yield in biculture. | species | Biomass density, biovol, etc. of a species $i$ interacting with another species $j$ (particular case of $Y_i$ for $S = 2$). | $B_i$ |
| Relative yield | species | $Y_i/K_i$ | $y_i$ |
| Relative yield in biculture | species | Relative yield of a species $i$ interacting with another species $j$, i.e. a special case of relative yield for $S = 2$. | $b_i$ |
| Competition coefficient of species $j$ on species $i$ | species | Measures the effect of species $j$ over the growth rate of species $i$. | $a_{ij}$ |
| Competition matrix | whole community | Matrix whose elements are the $a_{ij}$. | **A** |
| Mean competition coefficient | whole community | Mean of the interspecific ($i \neq j$) interaction coefficients $a_{ij}$. | $a$ |
| Relative yield total | whole community | The sum of the relative yields of the $S$ species building up the community. | $RYT$ |

## III. RESULTS AND DISCUSSION

It often occurs that only a subset $\mathcal{M}$ of the $S$ monoculture experiments and a subset $\mathcal{B}$ of the $S\times(S-1)/2$ biculture or pairwise experiments were carried out. The recipe we use for estimating the experimental average interspecific competition coefficient $a$ (to feed equation (9)) and the experimental $RYT$, which has demonstrated to work quite well, is as follows.

First, the average relative yields in biculture, $\langle b_i \rangle_{\mathcal{M}+\mathcal{B}}$, is computed from the measured yields of subsets $\mathcal{M}+\mathcal{B}$. Second, using eq. (10), $a_{\mathcal{M}+\mathcal{B}}$ can be approximated as:

$$a_{\mathcal{M}+\mathcal{B}} = (1-\langle b_i \rangle_{\mathcal{M}+\mathcal{B}})/\langle b_i \rangle_{\mathcal{M}+\mathcal{B}}. \tag{11}$$

Inserting $a_{\mathcal{M}+\mathcal{B}}$ into eq. (9) we get the theoretical corresponding $RYT$, $\widehat{RYT}(a_{\mathcal{M}+\mathcal{B}})$, to compare vs. the experimental $RYT_{\mathcal{M}}$ (the subscript $\mathcal{M}$ is to emphasize that this quantity is estimated from the subset of available monoculture experiments). Eq. (11) was also used in those experiments in which only mean yields for mono and bicultures were available and we approximated $\langle b \rangle$ by $\langle B \rangle/\langle K \rangle$ (see Table 1). There are at least three different ways to estimate $RYT_{\mathcal{M}}$. The first and straightforward procedure is by summing the available empirical relative yields $\{y_i\}_{\mathcal{M}}$. A second one is by approximating $RYT_{\mathcal{M}} = S\langle Y \rangle/\langle K \rangle_{\mathcal{M}}$ (e.g. Note [i] in Table 1). A third estimation, that only requires knowing the total yield of the $S$-species polyculture $\Sigma_i Y_i$, is $RYT_{\mathcal{M}} = \Sigma_i Y_i/\langle K \rangle_{\mathcal{M}}$ (e.g. Note [j] in Table 1). Box 2 illustrates this calculation of $a$ and how to estimate $RYT_{\mathcal{M}}$.

From Table 1 and Table S1 (which includes results for $S = 2$ species) we see that eq. (9) predicts the observed $RYT$ with a relative error $\varepsilon_{TE}$ of 1 % or less (10 % or less) for 35 % (76 %) of the experiments. This accuracy is remarkable since the experimental SE for such quantities are in general greater than $\varepsilon_{TE}$ (in the interval 2-43 %, mean = 21 % [28]). Therefore our formula allows predicting the $RYT$ simply from the knowledge of the mean empirical competition strength. Notice that as $S$ increases the fraction $f_e$ of the $S\times(S+1)/2$ experiments required to estimate the experimental quantities also decreases. Thus for $S > 4$, even when using $f_e \ll 100$ %, eq. (9) more frequently is able to predict the $RYT$ with a smaller error: for six experiments out of 14 the relative error is below 10 % (see Table 1).

Table 1. *RYT* predicted by (9) vs. experimental values for 25 ecological communities across different taxa (mainly plants, but also algae, insects and crustacean). $S > 2$ is the number of coexisting species; exp. and theo. denote, respectively, experimental and theoretical; $a$ is the measured mean interspecific competition coefficient; *RYT* is the relative yield total; $\varepsilon_{TE}$ is the relative difference of theo. *RYT* respect to exp. *RYT* as %; SE is the standard error as % and $f_e$ is the percentage of the total $S\times(S+1)/2$ experiments which were used to estimate the experimental quantities.

| Taxon | Community (site, year, treatment, etc) | S | exp. $a$ | exp. RYT | theo. RYT | $\varepsilon_{TE}$ % | exp. SE % | $f_e$ % | Ref. | Note |
|---|---|---|---|---|---|---|---|---|---|---|
| ALGAE | Santa Catalina Is., California | 3 | 0.61 | 1.53 | 1.35 | 12.4 | 14.1 | 100 | [13] | |
| PLANTS | Grassland, San Jose, Ca. 1998 | 3 | 0.23 | 1.54 | 2.06 | 33.4 | 27.6 | 100 | [14] | [a] |
| PLANTS | Iowa, 2003-2005, 1 cut | | | | | | | | [15] | [b] |
| | species mixture: IBF-IW-EG | 3 | 0.54 | 1.20 | 1.44 | 19.9 | 27.2 | 100 | | |
| | species mixture: IBF-SW-EG | 3 | 0.49 | 1.16 | 1.51 | 29.9 | 23.9 | 100 | | |
| ALGAE | continuous culture systems | 4 | 0.75 | 1.04 | 1.23 | 18.1 | >4.7 | 100 | [16] | |
| CRUSTACEAN | Lab. microcosms | 4 | 0.39 | 1.32 | 1.85 | 39.4 | >2.0 | 100 | [17] | [c] |
| PLANTS | Random distribution experim. | 4 | 0.98 | 0.92 | 1.02 | 11.0 | 18.1 | 100 | [18] | |
| PLANTS | Pastures, British Columbia | | | | | | | | [19] | |
| | 1939 pasture | 4 | 0.67 | 1.10 | 1.33 | 21.5 | 19.8 | 100 | | [d] |
| | 1958 pasture | 4 | 0.98 | 1.02 | 1.01 | 1.0 | 12.7 | 100 | | |
| | 1977 pasture | 4 | 0.91 | 1.01 | 1.07 | 6.6 | 14.0 | 100 | | |
| PROTOZOA | Cultures & lab. experiments | 4 | 0.62 | 1.08 | 1.62 | 49.6 | 43.5 | 100 | [20] | |
| PLANTS | Shoreline of Axe Lake, Ontario | 7 | 0.52 | 2.02 | 1.74 | 14.1 | NA | 100 | [21] | |
| PLANTS | Otago lawn, NZ | 7 | 0.75 | 1.26 | 1.24 | 1.6 | 4.5 | 100 | [22] | |
| PLANTS | BIODEPTH exp., year 3 | | | | | | | | [23] | |
| | Sweden, blok R6P009 | 7 | 0.12 | 5.05 | 4.56 | 9.6 | 40.3 | 9.5 | | [e] |
| | Sweden, blok R6P010 | 7 | 0.12 | 4.08 | 4.56 | 11.8 | 48.9 | 9.5 | | [e] |
| | Silwood, UK, mix.nest 233 | 11 | 0.46 | 2.04 | 1.97 | 3.3 | 13.2 | 10.9 | | [e],[f],[g] |
| | Silwood, UK, mix.nest 234 | 11 | 0.46 | 1.93 | 1.97 | 1.9 | 18.9 | 10.9 | | [e],[f],[g] |
| | Germany, mix.nest 28 | 12 | 0.55 | 2.15 | 1.71 | 20.1 | 29.9 | 7.6 | | [h],[i] |
| | Germany, mix.nest 29 | 12 | 0.55 | 1.73 | 1.71 | 0.8 | 14.9 | 7.6 | | [h],[i] |
| | Sheffield, UK, block S7P012 | 12 | 0.81 | 1.61 | 1.21 | 24.5 | 16.2 | 13.6 | | [h] |
| | Sheffield, UK, block S7P023 | 12 | 0.81 | 1.48 | 1.21 | 18.0 | 17.6 | 13.6 | | |
| | Sheffield, UK, block S7P023 | 12 | 0.81 | 1.61 | 1.21 | 24.8 | 16.2 | 13.6 | | |
| PLANTS | Cedar Creek E120, year 2003 | 16 | 0.24 | 4.28 | 3.73 | 12.9 | 5.9 | 10.4 | [24] | |
| PLANTS | Jena exp., year 2007 | 16 | 0.28 | 2.80 | 3.06 | 9.5 | 16.9 | 6.7 | [25] | [j] |
| PLANTS | Switzerland, BIODEPTH exp. y1-3 | 32 | 0.22 | 3.44 | 3.81 | 11.5 | 15.2 | 3.2 | [23] | |

[a] More aggregate data: for functional groups rather than individual species.
[b] Averages over two different places and across the three years.
[c] *K* obtained from the removal of 2 species rather than S-1 = 3.
[d] exp. *RYT* is a lower bound since there are other six additional species whose relative yields were not measured.
[e] The number of coexisting species *S* was smaller than the number of seeded species (12) in this plot.
[f] In mix.nest 233 (234) of Silwood all except one (all) of the 11 relative yields *y* of the polyculture are available.
[g] In spite of many positive interactions the % error is small.
[h] *a* was computed from mean biomass densities for mono & bicultures: $\langle b \rangle = \langle B \rangle_B /\langle K \rangle_M$, $a = (1-\langle b \rangle)/\langle b \rangle$.
[i] The experimental *RYT* was computed approximately as $S\langle Y\rangle/\langle K\rangle_M$.
[j] The experimental *RYT* was computed approximately by averaging $\sum_i Y_i/\langle K\rangle_M$.

A possible source explaining some of the $\varepsilon_{TE}$ errors is that our approach requires the coexistence of species in pairwise experiments in order to compute the corresponding competition coefficients by eq. (10). Obviously this procedure doesn't work if one of the species completely displaces the other. Some of the experimental data used in Table 1 include pairwise experiments in which the yield of one of the species was very small, *i.e.* the experiment was at the border between biculture and monoculture. An example is ref. [20], this could explain why it yields the largest $\varepsilon_{TE}$ (49.6 % as shown in Table 1). In order to improve accuracy, an alternative would be to discard those experiments and use an incomplete competition matrix and eq. (11).

**Box 2. An example showing both the calculation of theoretical *RYT* (eq. (9)) and how the corresponding experimental *RYT* can be estimated with incomplete data.**

> We use here the last experiment in Table 1, corresponding to the BIODEPTH experiments for grassland plant communities conducted in Switzerland [23], to illustrate how this table was built. Specifically we consider the annual measures of yield along three years[(*)] for 32-species polycultures.
>
> To completely estimate the LVC parameters 32 monoculture and 32×31/2= 496 biculture experiments would be required. Of course, for practical reasons, only few of these experiments were carried out: 10 monoculture and 7 biculture experiments, i.e. a fraction of the total = (10+7)/(32+496) or a $f_e$ = 3.2 %. The $\langle b_i \rangle_{M+B}$ that emerges from the 17 monoculture and biculture experiments is 0.74 [23], and therefore, by equation (11), $a_{M+B}$ = 0.26/0.74 = 0.35. Substituting in equation (9) we obtain a theoretical *RYT*, $\widehat{RYT}(a_{M+B})$ = 32/(31×0.35+1) = 2.70.
>
> As mentioned, there are three different ways to estimate $RYT_M$. The first one is by summing the 10 relative yields obtained from the 10 monoculture experiments and the 32-species polyculture experiment and we obtain $RYT_M$ = 3.05 [23]. The second estimation is from the mean monoculture yields we have: $\langle K \rangle_M$ = 275 g/m² and $\langle Y \rangle$ = 27g/m² [23], which produces $RYT_M = S\langle Y \rangle/\langle K \rangle_M$ = 32×27/275 = 3.14. The third estimate for $RYT_M$ is by taking the ratio between the total yield in the 32-species polyculture $\sum_i Y_i$ = 661 g/ m² [23] and $\langle K \rangle_M$: $RYT_M$ = $\sum_i Y_i/\langle K \rangle_M$ = 661/275 = 2.40. This three estimates are not dramatically different and the relative errors between the theoretical and experimental *RYT* they yield are, respectively, $\varepsilon_{TE}$ = 11.5 %, 14.0 % and 12.2 % which are all smaller than the experimental standard error = 15.2 %. Therefore the error calculation seems to be reliable.
>
> (*): Since $f_e$ was so low we consider the three years rather than only the last year to have more statistics.

Fig. 1 shows he *RYT* measured in these 77 experiments as well as their theoretical predictions by eq. (9) as a function of both *S* and *a*. It is known that the mean field approximation becomes exact in the (artificial) limit $S \to \infty$ [29]. Thus one expects that eq. (9) becomes better the larger the number of interacting species $S$. However, we observe that eq. (9) reproduces with great accuracy also the experimental $RYT$ for $S = 2$. This good agreement for $S = 2$ can be understood because the quantities used for computing $a$ and the $RYT$ are the same: $B_i$. In fact the difference can be quantified as follows. For $S = 2$, the approximated $RYT$ is given by:

$$\widehat{RYT}_{S=2} = 2/(1+a). \tag{9'}$$

To compare (9') with its exact value, solving (3) we obtain the biculture ($S = 2$) relative yields:

$$b_i = \frac{1-a_{ij}}{1-a_{ij}a_{ji}} \quad \text{and} \quad b_j = \frac{1-a_{ji}}{1-a_{ij}a_{ji}}, \tag{12}$$

in such a way that the $RYT$ is given by:

$$RYT_{S=2} = b_i + b_j = 2 \bigg/ \left(1+\langle a_{ij}\rangle_{i \neq j} + \delta\right) \quad \text{with} \quad \delta = \frac{(a_{ij}-a_{ji})^2}{4\left(1-\langle a_{ij}\rangle_{i \neq j}\right)}. \tag{13}$$

By taking $a = \langle a_{ij}\rangle_{i \neq j}$, (13) reduces to (9') whenever $\delta$ is negligible compared with $1+\langle a_{ij}\rangle_{i \neq j}$. The points that depart most from the curve defined by (9') are those for which $\delta$ of eq. (13) is relatively large (*e.g.* the lowest point at $a = 0.5$, $S = 2$ in Fig. 1), so that the approximation (9') breaks down (see Table S1).

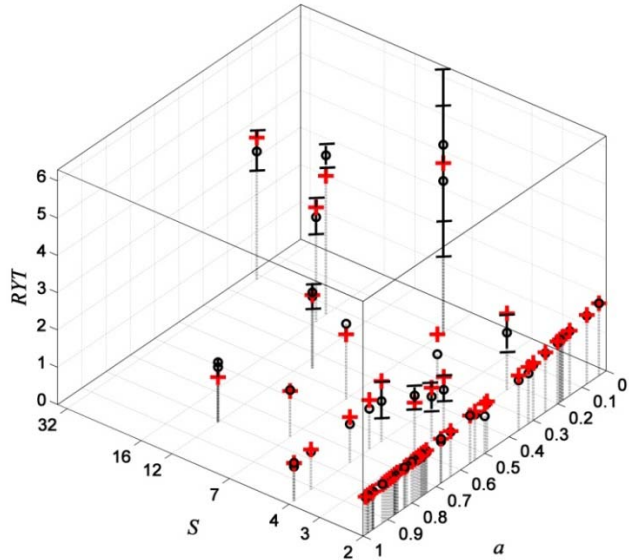

Figure 1. Theoretical $RYT$ predicted by eq. (9) (red crosses) vs. empirical $RYT$ (o) with error bars corresponding to ± SE for the 77 experiments as a function of $S$ (log scale) and $a$. (See Table 1 for the size of the standard experimental errors for $S > 2$.)

Notice that the *RYT* increases with *S*, a phenomenon which has been observed in many empirical studies and interpreted as niche partitioning [30]. That is, different species can use resources in a complementary way and therefore more diverse communities will lead to a higher *RYT*. However it was noticed that in spite of widespread claims about the role of biodiversity and ecosystem functioning relationships, researchers have rarely provided any direct quantitative evidence that niche partitioning is responsible for higher productivity in mixtures. Rather many claims about niche partitioning come from qualitative interpretations of species natural histories or differences in functional traits [31]. The *RYT* > 1 for the great majority of the experiments considered here, especially for those with *S* > 2 interacting species (24 out of 25 as shown in Table 1), is a clear *quantitative* evidence of such species complementarity [7,8]. Moreover, we are providing the functional dependency for *RYT* with *S*: a function which increases with *S* more slowly than linearly. As far as we know this is a novel finding.

Fig. 2 shows the *MRY* = *RYT*/*S*. As expected,, we see that the *MRY* for fixed *S* decreases with *a*, while for fixed *a* the *MRY* also decreases with *S*.

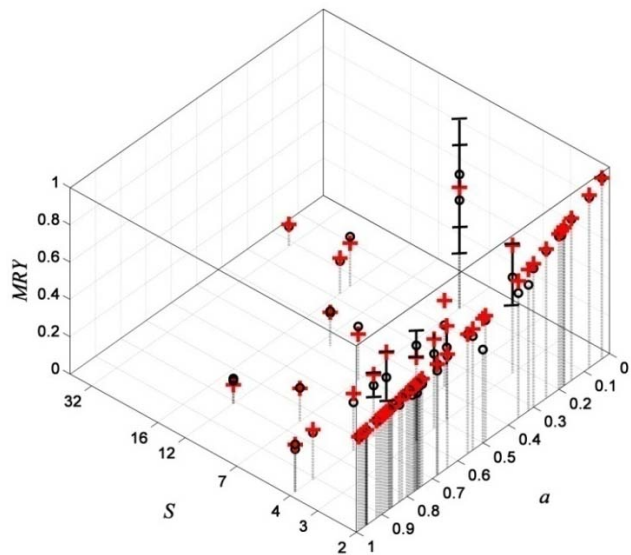

Figure 2. Theoretical *MRY* predicted by eq. (10) (red crosses) vs. empirical *MRY* (o) with error bars corresponding to ± SE for the 77 experiments as a function of *S*(log scale) and *a*.

We end this section with three comments/remarks about the method we are proposing.

- First, our goal in designing the present framework is, given a stable community at the equilibrium point, to explain *ex post* the productivity by using the mean interaction parameter estimated from the pairwise experiments.

- The former point brings us to the issue regarding the robustness of our framework: it can happen that the interaction matrix **A** estimated from all pairwise experiments leads to an equilibrium state in which not all the species coexist. That is, the solution to eq. 3 (eq. 5) is a fixed point of the system, but this fixed point can be unstable or unfeasible (*i.e.* involving negative species densities). In the latter case, eq. 3 does not hold for the full set of species. Instead, the density of some of the species is zero at equilibrium. However, this does not represent a serious problem for our method since it does not rely on "microscopic" information like the detailed equilibrium state. We can always use a subset of the competition coefficients in such a way that the equilibrium state remains undefined and this method basically still provides quite accurate predictions (as shown in Table 1). Therefore the proposed approximation provides a good estimate even in cases where, mathematically, not all species survive and eq. 3 does not strictly apply. At any rate, if we insist on working with an entire interaction matrix that leads to a mathematical equilibrium which is unstable/unfeasible, there is a way to circumvent this problem. We still have the freedom to modify the parameters around the computed mean values, provided we keep them within their confidence intervals, in such a way to attempt to transform the unstable equilibrium into a stable one. This solution was shown to work for grassland experiments [26, 27].

- Third, strictly speaking our framework cannot be used to predict the productivity of a combination of species *ex ante*. This is because it would not be possible to predict if this particular species combination was stable and feasible based only on the mean interaction parameter; for doing this we need the entire interaction matrix **A** to compute the equilibrium state through eq. (5). However, this framework can be used to provide an estimate of what would be the productivity of an artificially designed community in the case that it is stable and feasible. Or, it can also be used to predict the impact on a given community of species loss affecting both $S$ and $a$.

## IV. CONCLUSION

The observed complexity of interspecific interactions often leads to distrust of highly simplified mathematical equations like LVCT and their ability to make quantitative predictions about community structure. More complex theories may be preferred, perceived as more reliable, although they can be as intractable as the real systems they aim to model. We prefer appealing to the smallest possible number of parameters and, more importantly, to parameters that can be practically estimated from empirical data, either experiments or field work. Therefore here we contrasted predictions for global observables of the simplest possible theory for a community of competing species, LVCT, against empirical results for a wide variety of communities (from protozoa to mammals). We showed that LVCT works pretty well as a quantitative tool for predicting *MRY* and *RYT* from the knowledge of the mean interaction intensity between pairs of species, both for $S = 2$ and large values of $S$ (independently of the taxon).We are not denying the existence of higher order interactions (non-linear terms) in some communities [32], but our analysis seems to confirm that they are more the exception than the rule [33]. These higher order interactions would lead to additional competition parameters for such organisms having different competitive effects when acting in coalitions and when acting alone. A cited example is the microcrustacean community of ref. [17], which would explain why our approximation leads to such a large relative error for this experiment (39 %) as shown in Table 1).

To conclude, our point is that parsimonious modelling is particularly compelling for developing quantitatively predictive tools for community ecology. This is because mathematical treatment and, more importantly, adequate experimental design and field data would become much more difficult with such higher order interactions. Besides, the experimental SE are in general large. This precludes falsification of such more elaborate models which should be considered only for those particular communities in which the simpler LVCT fails. The kind of approximation we consider here would be similar to the mean-field van der Waals equation of state which is not exactly true for any real gas, but it frequently provides a description that is good enough for many practical purposes and is proportional to the problem that it seeks to solve. To explore the limits the applicability of the method it seems interesting to study how it performs in different situations when relaxing the conditions of equilibrium and of interactions dominated by competition. Work is in progress towards this direction.

ACKNOWLEDGMENTS
The author thanks Cedar Creek LTER team, S. Roxburgh and A. Segura for sharing experimental data. Work supported in part by ANII-Uruguay (SNI and project ERANET-LAC).